# If your *P* value looks too good to be true, it probably is: Communicating reproducibility and variability in cell biology

Samuel J. Lord,[a] Katrina B. Velle,[b] R. Dyche Mullins,[a*] Lillian K. Fritz-Laylin[b*]

[a] Department of Cellular and Molecular Pharmacology, Howard Hughes Medical Institute, University of California, San Francisco CA 94143
[b] Department of Biology, University of Massachusetts, Amherst MA 01003
* Corresponding authors: dyche.mullins@ucsf.edu & lfritzlaylin@umass.edu

ABSTRACT: The cell biology literature is littered with erroneously tiny *P* values, often the result of evaluating individual cells as independent samples. Because readers use *P* values and error bars to infer whether a reported difference would likely recur if the experiment were repeated, the sample size *N* used for statistical tests should actually be the number of times an experiment is performed, not the number of cells (or subcellular structures) analyzed across all experiments. *P* values calculated using the number of cells do not reflect the reproducibility of the result and are thus highly misleading. To help authors avoid this mistake, we provide examples and practical tutorials for creating figures that communicate both the cell-level variability and the experimental reproducibility.

SUMMARY SENTENCE: The sample size for a typical cell biology experiment is the number of times the experiment was repeated, not the number of cells measured.

## *P* = 0.00000000000000001?! Think again

Error bars and *P* values are often used to assure readers of a real and persistent difference between populations or treatments. *P* values are based on the difference between population means (or other summary metric) as well as the number of measurements used to determine that difference. In general, increasing the number of measurements decreases the resulting *P* value. To convey anything about experimental reproducibility, *P* values and standard error of the mean should be calculated using independent measurements of a population of interest, typically observations from independent samples or separate experiments (AKA "biological replicates") (Lazic 2010; Naegle et al. 2015; Lazic et al. 2018; Aarts et al. 2015). Limited time and resources usually constrain cell biologists to repeat any particular experiment only a handful of times, so a typical sample size *N* is something like 4. Too often, however, authors[1] mistakenly assign *N* as the number of cells, or even the number of subcellular structures, observed during the experiment. This number may be on the order of hundreds or thousands, resulting in vanishingly small *P* values and tiny error bars that are not useful for determining the reproducibility of the experiment.[2]

Cell biology data is inherently noisy. Compounding this variability, quantitative experiments are difficult to replicate exactly from day-to-day and lab-to-lab. Well-designed studies embrace both cell-to-cell and sample-to-sample variation (Altman and Krzywinski 2015). Unlike measuring the gravitational constant or the speed of light, repeatedly quantifying a biological parameter rarely converges on a single "true" value, due to the complexity of living cells or because many biological processes are intrinsically stochastic (Raj and van Oudenaarden 2008). Calculating standard error from thousands of cells conceals this expected variability. We have written this tutorial to help cell biologists calculate meaningful *P* values and plot data to highlight both experimental robustness *and* cell-to-cell variability. Specifically, we propose the use of distribution–reproducibility "SuperPlots" that display the distribution of the entire dataset, and report statistics (such as means, error bars, and *P* values) that address the reproducibility of the findings.

---

[1] We freely admit that our past selves are not innocent of the mistakes described in this manuscript.
[2] The mistake of treating dependent measurements as separate samples is sometimes referred to as "pseudoreplication" (Hurlbert 1984).

While far from perfect,[3] the *P* value does offer a pragmatic metric to infer whether an observed difference is reproducible and substantial relative to the noise in the measurements (Greenwald et al. 1996). And the problem of inflated *N* would only migrate with us if the field transitioned to confidence intervals or estimation statistics, given that confidence intervals are calculated with the same parameters as the t-test (Altman and Bland 2011). This is not a tutorial on statistical analysis; many have described the troubles with *P* values (Sullivan and Feinn 2012; Greenland et al. 2016; Gardner and Altman 1986) and there are several excellent practical guides to statistics for cell biologists (Pollard et al. 2019; Lamb et al. 2008). In this paper we specifically address simple ways to communicate reproducibility when performing statistical tests and plotting data.

**What Population is being Sampled?**

Here's a simple question to help you clarify what your sample size *N* should be: *What population are you trying to sample?* The choice of *N* determines the population that being evaluated or compared (Naegle et al. 2015; Lazic et al. 2018; Pollard et al. 2019). For instance, if I measure hair length from many different people (*N* = number of people), I have evaluated a subset of the statistical population "humans." If I measure the hair from many people, dogs, cats, elephants, whales, and squirrels, then I am sampling mammals. If I measure many of my own hairs (*N* = number of hairs), then the population sampled is my own head, and it would be unreasonable to make inferences about the hair length of all humans or all mammals (Vaux et al. 2012).

A typical cell biology experiment strives to draw general conclusions about all similar cells, so the sample selection should reflect that broad population. For example, if you want to know if a particular treatment changes the speed of crawling cells, you could split a flask of lymphocytes into two wells of a 96-well plate, dose one well with a drug of interest and one with a placebo, and then track individual cells in each of the two wells. If you use each cell as a sample (*N* = number of cells), the two populations you end up comparing are the cells in *those two particular wells*. By repeating the experiment multiple times from new flasks, and using each experiment as a sample (*N* = number of independent experiments), you evaluate the effect of the treatment on any arbitrary flask of similar cells. Multiple observations within one well increases the precision for estimating the mean for that one sample, but doesn't reveal a truth about all cells in all wells (just like measuring many hairs on my own head doesn't give me insight into the average haircut).

If you only care about cell-to-cell variability within a particular sample, then maybe *N* really is the number of cells you observed. Making inferences beyond that sample, however, would be questionable, because the natural variability of individual cells can be overshadowed by systematic differences between biological replicates. Whether caused by passage number, confluency, or location in the incubator, cells often vary from sample-to-sample and day-to-day. Entire flasks of cells might even be described as "unhappy." Accordingly, cells from experimental and control samples (e.g. tubes, flasks, wells, coverslips, rats, tissue samples, etc.) may differ from each other, regardless of the intended experimental treatment. When authors report the sample size as the number of cells, the resulting statistical analysis cannot help the reader evaluate whether any observed differences are due to the intended treatment or simple sample-to-sample variability. We are not prescribing any specific definition of *N*, we are simply encouraging researchers to consider what main source of variability they hope to overcome when designing experiments and statistical analyses (Altman and Krzywinski 2015) (see Table 1).

**Statistics in Cell Biology Typically Assume Independent Tests of a Hypothesis**

To test the hypothesis that two treatments or populations are different, the treatment must be applied or the populations sampled multiple times. In a drug trial for a new painkiller pill, one of my knees cannot be randomly assigned to the treatment group and the other to placebo, so researchers cannot count each of my knees as a separate *N*. (If the trial is testing steroid injections, then under certain statistical models (Aarts et al. 2015), each knee could be considered a separate sample.) Similarly, neighboring cells within one flask or well treated with a drug are not separate tests of the hypothesis, because the treatment was only applied once. But if individual cells are microinjected with a drug or otherwise randomly assigned to a different treatment, then each cell really can be a separate test of a hypothesis.

---

[3] There is nothing magic about 0.05.

Finding truly independent groups and deciding what makes for a good biological replicate can be challenging (Naegle et al. 2015; Blainey et al. 2014; Aarts et al. 2015). For example, is it acceptable to run multiple experiments from just one thawed aliquot of cells, or do I need to borrow an aliquot from another lab? Is it necessary to generate multiple knockout strains? Is it sufficient to test in one cell line, or do I need to use multiple cell types or even cells from multiple species? Can I perform all experiments in one lab, or should I include results from a lab on the other side of the country (Lithgow et al. 2017)? There's no single right answer: each researcher must balance practicality with robust experimental design. At a minimum, researchers must perform an experiment multiple times if they want to know whether the results are robust.[4]

**How to Calculate *P* Values from Cell-Level Observations**

Cell biologists often observe hundreds of cells per experiment and repeat an experiment multiple times. To leverage that work into robust statistics, one needs to take into account the hierarchy of the data. Simply combining all the cell-level data from multiple independent experiments squanders the available information about run-to-run variability (Figure 1). There is ample literature about the analysis of this type of hierarchical data (Galbraith et al. 2010; Gelman and Hill 2006), which takes into account both the variance within a sample as well as the clustering across multiple experimental runs (Aarts et al. 2015), or that propagate the error up the chain, such as a nested ANOVA (Krzywinski et al. 2014).

Given the problems with null-hypothesis significance testing, it may be better to avoid *P* values altogether. Analyses that use effect sizes (Sullivan and Feinn 2012) and confidence intervals (Gardner and Altman 1986) summarize differences between treatments without defining an arbitrary measure of "significance." Meta-analyses of multiple clinical trials often use estimation statistics and summarize multiple studies together into one "forest" plot (Lewis and Clarke 2001), which includes a summary value resulting from the meta-analysis (Deeks et al. 2001). Cell biologists could emulate clinical researchers, evaluating each day's experiment as if it were a separate trial. But there are other ways to quickly and easily communicate run-to-run variability.

A simple approach—which permits conventional t-test or ANOVA calculations—is to average the cell-level data for each separate experiment and compare the subsequent sample-level means (Galbraith et al. 2010; Altman and Bland 1997; Lazic 2010). For example, if you have three biological replicates of control and treated samples, and you measure the cell diameter of 200 cells in each sample, first calculate the mean of those 200 measurements for each sample, then run a t-test on those sample means (three control, three treated). This simple approach might fail to detect small but real differences between groups,[5] where more advanced techniques may prove to be more powerful (Galbraith et al. 2010; Aarts et al. 2015). Another way this simple approach can falter is if a different number of cells are observed in each round, or there is drastically different variance in each sample, in which case it would be better to weight each sample by its precision (Deeks et al. 2001). Nevertheless, although better analyses exist, averaging dependent observations together is simple and avoids false positives (Galbraith et al. 2010; Aarts et al. 2015).

**Communicating Variability with SuperPlots**

After analyzing hundreds of cells across multiple rounds of experimentation (and calculating useful *P*

---

[4] This raises the question of how many cells one should look at in each sample. Is it better to look at many cells in a few biological replicates or spend less time measuring individual cells and redirect that effort to repeating the experiment additional times? Multiple analyses have found that increasing the number of biological replicates usually has a larger influence on the statistical power than imaging many more cells in each sample (Blainey et al. 2014; Aarts et al. 2015).

[5] Fortunately, the loss in power is typically small (Aarts et al. 2015).

values), it would be nice to incorporate both the cell-level variability and experimental repeatability of all that work into a single diagram. Figure 1 shows plots that can be improved with correct thinking about independent samples.

If we summarized all (human) hairstyles with a single average hair length, it would not describe asymmetrical "hipster" haircuts nor capture the large range of hair lengths.[6] Similarly, bar graphs are problematic because they obscure the distribution of cell-level data as well as the sample-to-sample repeatability (Weissgerber et al. 2015). Although it's tempting to place the blame on the use of bar graphs, the problem is deeper. In Figure 1, the plots on the left have small error bars and comically tiny *P* values, which should raise red flags given how difficult it would be to replicate a cell biology experiment with identical results and/or to repeat it hundreds of times, which such miniscule *P* values imply. While beeswarm, column scatter, box-and-whisker, and violin plots are great at conveying information about the range and distribution of the underlying data, plotting the entire dataset does not make it appropriate to treat repeated measurements on the same sample as independent experiments.

Therefore, we suggest authors incorporate information about distribution and reproducibility by creating "SuperPlots," which superimpose summary statistics from repeated experiments on a graph of the entire cell-level dataset (Figure 1, right columns). Not only does a SuperPlot convey more information than a conventional bar graph or beeswarm plot, it also makes it clear that statistical analyses (e.g. error bars and *P* values) are correctly calculated across separate experiments, not individual cells—even when each cell is represented on the plot. For example, the mean from each experiment could be listed in the caption, or plotted as a larger dot on top of the many smaller dots that denote individual measurements.

When possible, it is best to link samples by run, for instance, by color-coding the dots by experiment or drawing a line linking paired measurements together (Figure S1D). The benefit of these linkages is to convey the repeatability of the work: readers learn more if they know that one experiment exhibited high readings across the board than if they have to guess the trend in each sample (Figure 1 B vs C, right columns). An additional benefit to linking data is that it eliminates the need to normalize data in order to directly compare different experimental runs. Often, multiple experiments might all exhibit the same *trend*, but different absolute values (Figure 1B). By encoding the biological replicate into the data, such trends can be revealed without normalizing to a control group. *P* values can then be calculated using statistical tests that take into account linkages among samples (e.g. a paired or ratio t test).[7]

An impressive amount of information can be crammed into color-coded beeswarm SuperPlots (see Figure 1, rightmost plots), where each cell-level datapoint divulges which experiment it came from (Galbraith et al. 2010; Weissgerber et al. 2017). This helps convey to the reader whether each experimental round gave similar results or if one run biases the conclusion (Figure 1C). The summary statistics and *P* values in beeswarm SuperPlots (calculated using the means from each experiment, not the value from each individual cell) are overlaid on the color-coded scatter. (See Figures S2-5 for tutorials on how to make beeswarm SuperPlots in Prism, Python, R, and Excel).

However an author chooses to display their data, it is critical to list the number of independent experiments in the figure or caption. For SuperPlots, the caption should explain that the means and the statistical tests (e.g. t-tests, ANOVAs, and post tests) were calculated on the average observed value for each independent experiment.

**Error Bars that Communicate Reproducibility**

The choice of error bars on a SuperPlot depends on what you hope to communicate: *descriptive* error bars characterize the distribution of measurements (e.g. standard deviation), while *inferential* error bars evaluate how likely it is that the same result would occur if the experiment were to be repeated (e.g. standard error of the mean or confidence intervals) (Cumming et al. 2007). To convey how repeatable an experiment is, it is appropriate to choose inferential error bars calculated using the number of independent experiments as the sample size. However, calculating standard error of the mean by inputting data from all cells individually fails in two ways: first, the natural variability we expect from biology would be better summarized with a descriptive measure, like standard deviation; and second, the inflated *N* produces error bars that are artificially small (because the calculation for standard error of the mean includes $\sqrt{N}$ in the denominator) and communicate nothing about the repeatability of the experiment. Just like measuring my own hair many times gives me a precise estimate of my own hair length, but cannot be used to infer the average haircut.

---

[6] In this metaphor, a bar graph would imply that everyone has a bowl cut.
[7] In fact, not taking into account linkages can make the t-test too conservative, yielding false *negatives* (Galbraith et al. 2010).

The problems that arise from calculating error bars using cell number as the sample size are illustrated by comparing the left-hand panels of Figure 1A-B: when each cell measurement is erroneously treated as an independent replicate, the standard error of the mean is equally tiny in plots A and B, despite high variability between experimental replicates in B. In contrast, the SuperPlots in the right-hand columns show error bars that were calculated using biological replicates. Note that the standard error of the mean grows when the results vary day-to-day (Figure 1B, right columns). In cases where displaying every data point is not practical, authors should consider some way of representing the cell-to-cell variability as well as the run-to-run repeatability. This could be error bars that represent the standard deviation of the entire dataset, but with *P* values calculated from biological replicates.

**Conclusion**

Despite our emphasis on replication, a scientist can gather a lot of information by observing a single cell. In fact, some of the greatest discoveries had an *N* of one (Marshall et al. 1985). The field will squander opportunities for conceptual breakthroughs if it leaves no room for fleeting observations and singular results (Abercrombie et al. 1970; Tilney and Portnoy 1989). Such striking isolated observations, however, should not be garnished with meaningless *P* values.

On a practical note, if you calculate a *P* value smaller than Planck's constant, take a moment to consider: What variability does your *P* value represent? How many *independent* experiments have you performed, and does this match with your *N*? (See Table 1 for practical examples of this analysis.) Reviewers and editors should gently remind authors to state the *N* in terms of independent measurements and calculate *P* values based on that, rather than the total number of cells observed in each sample. We encourage authors and editors to focus not on reporting satisfying yet superficial statistical tests such as *P* values, but instead on presenting the data in a manner that conveys both the variability and the *reproducibility* of the work.

**Acknowledgements**

We are grateful to several colleagues who read this manuscript and provided useful feedback, including Kenneth Campellone, Adam Zweifach, William Velle, Geoff O'Donoghue, and Nico Stuurman. We also thank Natalie Petek for providing some hints on using Prism and Jiongyi Tan for reviewing the Python code. This work was supported by grants to LKF-L from the National Institutes of Health (grant 1R21AI139363 from the National Institute of Allergy and Infectious Diseases), by the National Science Foundation (grant IOS-1827257), and from the Pew Scholars Program in the Biomedical Sciences; and by grants to RDM from the National Institutes of Health (R35-GM118119) and Howard Hughes Medical Institute.

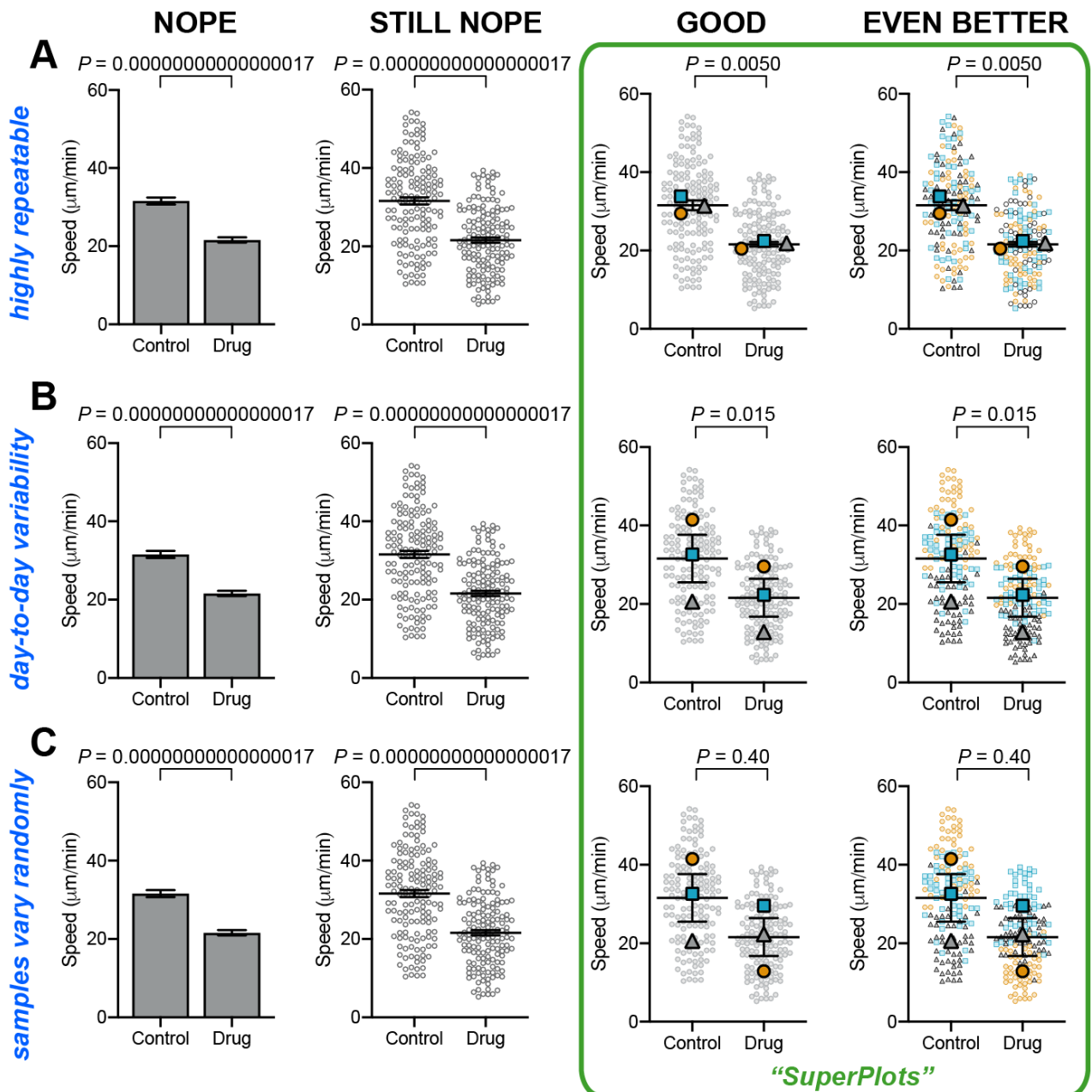

**Figure 1. The importance of displaying reproducibility**
The problematic figures in the "Nope" and "Still Nope" columns on the left treat *N* as the number of cells, resulting in tiny error bars and comically small *P* values. Also, these plots conceal any systematic error run-to-run, convolving it with cell-to-cell variability. To correct that, in the "Good" and "Even Better" columns, "SuperPlots" superimpose summary statistics from independent experiments (AKA "biological replicates") on top of data from all cells, and *P* values were correctly calculated using an *N* of three, not 300. In this case, the cell-level values were first pooled for each biological replicate and the mean calculated of each pool; those three means were then used to calculate the average (horizontal bar), standard error of the mean (error bars), and *P* value. Furthermore, each biological replicate is color-coded: the averages from one experimental run are yellow dots, another independent experiment is represented by gray triangles, and a third experiment is shown as blue squares. This helps convey that the trend is observed within each experimental run, as well as for the dataset as a whole. The beeswarm SuperPlots in the rightmost column represent each cell with a dot that is color coded according to the biological replicate it came from. **(A)** shows an example with highly repeatable data, **(B)** shows day-to-day variability, but a consistent trend, and **(C)** is dominated by one random run. Note that the plots that treat each cell as its own *N* (left columns) utterly fail to distinguish the three scenarios, claiming a significant difference after drug treatment, even when that is the result of random fluctuations. The *P* values represent an unpaired two-tailed t test for the left columns and a paired two-tailed t test for the right columns. For tutorials on making SuperPlots, see the supporting information.

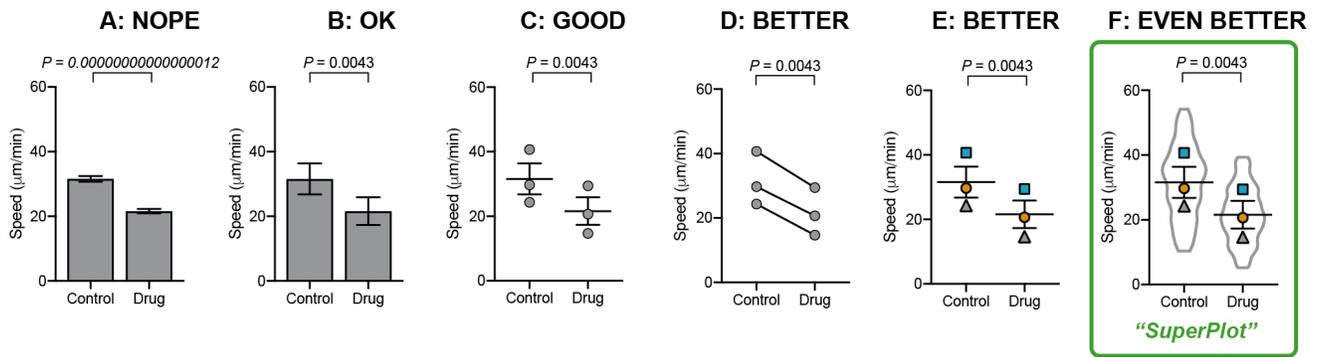

**Figure S1. Other plotting examples**
Bar plots can be enhanced even without using beeswarm plots. (**A**) The bar plot that calculates *P* and error bars using the number of cells as *N*. (**B**) But a bar graph can be corrected by plotting the mean of the replicates. (**C**) Showing each replicate reveals more than a simple bar graph. (**D-E**) Linking each pair by the replicate conveys important information about the trend in each experiment. (**F**) A SuperPlot shows not only information about each replicate and the trends, but also superimposes the distribution of the cell-level data, here using a violin plot.

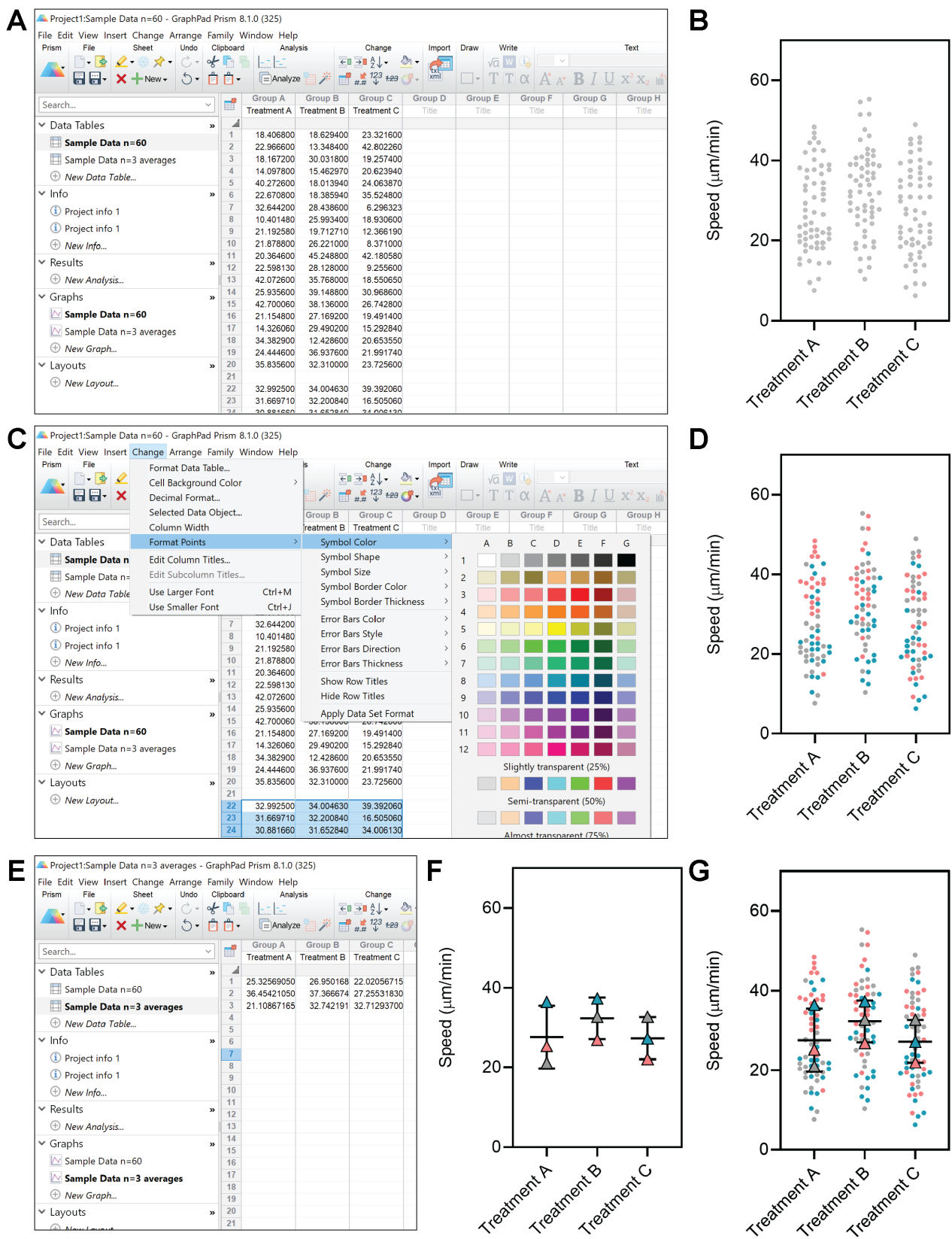

**Figure S2: Tutorial for Making SuperPlots in Prism**
We describe how to make SuperPlots in GraphPad Prism 8 graphing software. If using other graphing software, one may create a separate, different colored plot for each replicate, then overlay those plots in software like Adobe Illustrator. (**A**) When adding data to the table, leave a blank row between replicates.

(**B**) Create a new graph of this existing data; under type of graph select "Column" and "Individual values," and select "No line or error bar." (**C**) After formatting the universal features of plot from B (e.g. symbol size, font, axes), go back to the data table and highlight the data values that correspond to one of the replicates. Under the "Change" menu, select "Format Points" and change the color, shape, etc. of the subset of points that correspond to that replicate. (**D**) Repeat for the other replicates to produce a graph with each trial color coded. (**E-F**) To display summary statistics, take the average of the technical replicates in each biological replicate (so you will have one value for each condition from each biological replicate), and enter those averages into another data table and graph. Use this data sheet that contains only the averages to run statistical tests. (**G**) To make a plot that combines the full dataset with the correct summary statistics, format this graph and overlay it with the above scatter SuperPlots (In Prism, this can be done on a "Layout."). This process could be tweaked to display other overlayed, color-coded plots (e.g. violin).

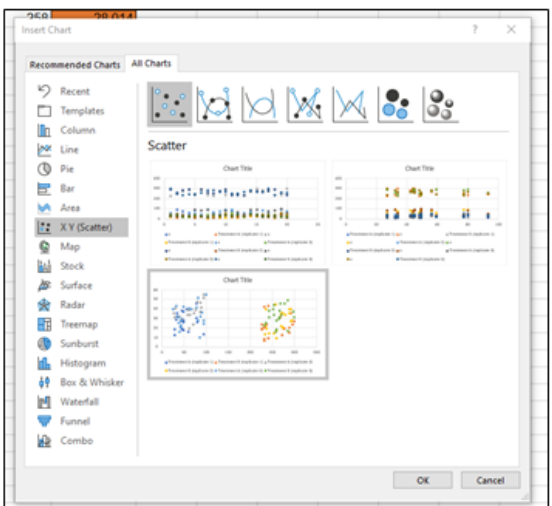
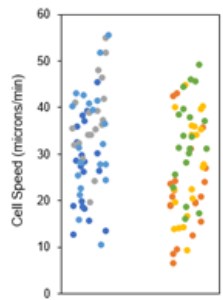
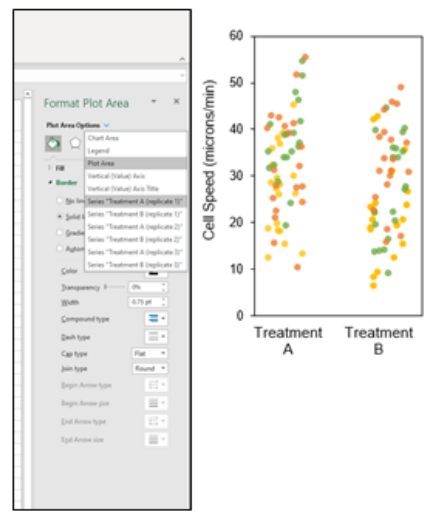
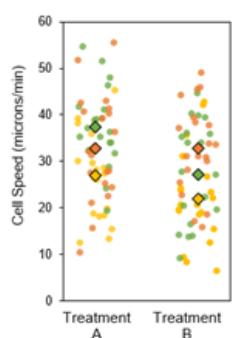
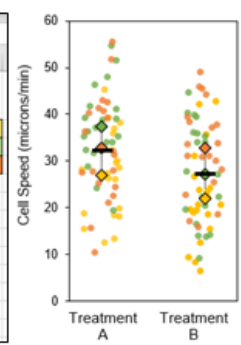

**Figure S3: Tutorial for Making SuperPlots in Excel**
(**A**) To make a SuperPlot using Excel (Microsoft Office 365 ProPlus for Windows; version 1912; Build 12325.20172), enter the values for the first replicate for the first condition into column B (highlighted in yellow), the second condition into column D (highlighted in yellow), and continue to skip columns between data sets for the remaining conditions and replicates (in this example, replicate 2 is highlighted in green and replicate 3 is in orange). For example, "Treatment A" could be control cells and "Treatment B" could be drug-treated cells. Label the empty columns as "x" and, starting with column A, enter random values to generate the scatter effect by using the formula "=RANDBETWEEN(25, 100)". To create a gap between the data sets A and B, use larger X values for treatment B by entering the formula "=RANDBETWEEN(225, 300)". (**B**) Highlight all the data and headings. In the insert menu, expand the charts menu to open the "Insert Chart" dialog box. Select "All Charts," and choose "X Y Scatter." Select the option that has Y values corresponding to your data sets. (In Excel for Mac, there is not a separate dialog box. Instead, make a scatter plot, right click on the plot and select "Select Data," remove the "x" columns from the list, then manually select the corresponding "X values =" for each dataset.) (**C**) Change the general properties of the graph to your liking. In this example, we removed the chart title and the gridlines, added a black outline to the chart area, resized the graph, adjusted the X axis range to 0-325, removed the X axis labels, added a Y axis title and tick marks, changed the font to Arial, and changed the font color to black. This style can be saved as a template for future use by right clicking. We recommend keeping the figure legend until the next step. (**D**) Next, double click the graph to open the "Format Plot Area" panel. Under "Chart Options," select your first data set, "Series 'Treatment A (replicate 1)." (On a Mac, click on a datapoint from one of the replicates, right click and select "Format Data Series.") Select "Marker" and change the color and style of the data points. Repeat with the remaining data sets so that the colors, shapes, etc. correspond to the biological replicate the data points came from. Delete the chart legend and add axis labels with the text tool if desired. (**E**) Calculate the average for each replicate for each condition, and pair this value with the X coordinate of 62.5 for the first treatment, and 262.5 for the second treatment to center the values in the scatterplot. Then, click the graph, and under the "Chart Design" menu, click "Select Data." Under "Legend Entries (Series)," select "Add" and under series name, select the three trial names, then select all three X and Y values for first treatment condition for "Series X Values" and "Series Y Values," respectively. Repeat for the second treatment condition, and hit "OK." (**F**) On the chart, select the data point corresponding to the first average and double click to isolate the data point. Format the size, color, etc. and repeat for remaining data points. (**G**) Optional: to add an average and error bars, either generate a second graph and overlay the data, or calculate the average and standard deviation using excel and add the data series to the graph as was done in E-F, using the "-" symbol for the data point.

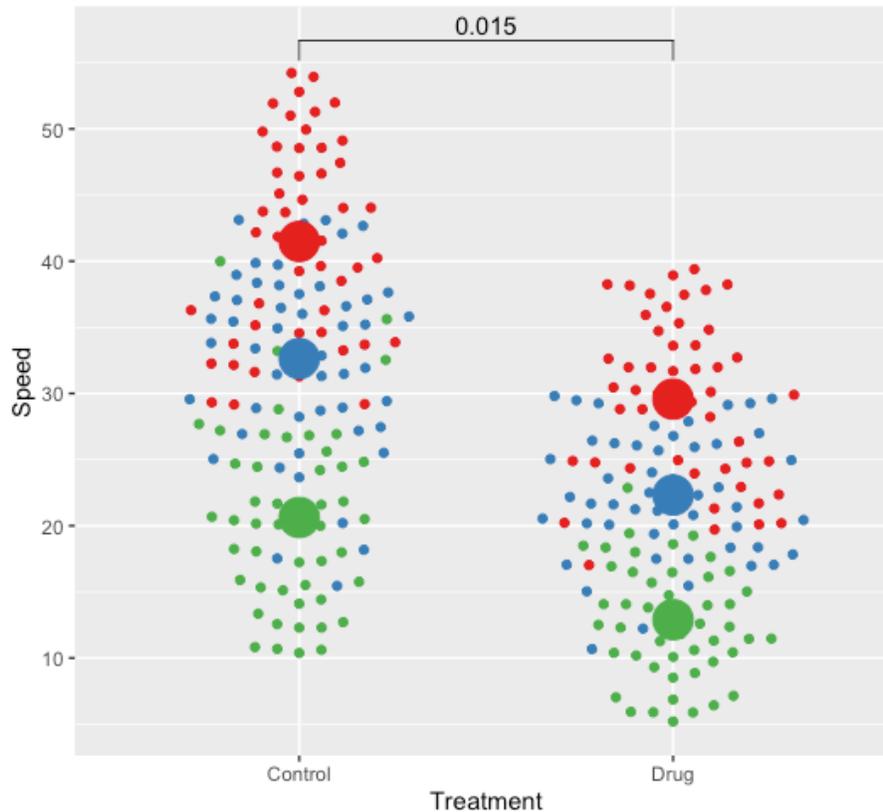

**Figure S4: Tutorial for Making SuperPlots in R**
Here is some simple code to help make SuperPlots in R using the ggplot2, ggpubr, dplyr, and ggbeeswarm packages. Dataset "combined.csv" is included in the supporting information.

```
install.packages("ggplot2")
install.packages("ggpubr")
install.packages("ggbeeswarm")
install.packages("dplyr")

library(ggplot2)
library(ggpubr)
library(ggbeeswarm)
library(dplyr)

#import dataset called "combined" that has the columns "Replicate," "Treatment," and "Speed."
      Change these and the code below if necessary.

#calculating averages of each replicate
ReplicateAverages <- combined %>% group_by(Treatment, Replicate) %>%
      summarise_each(list(mean))

#plot Superplot and P value (paired t test) based on biological replicate averages

ggplot(combined, aes(x=Treatment,y=Speed,color=factor(Replicate))) + geom_beeswarm(cex=3) +
      scale_colour_brewer(palette = "Set1") + geom_beeswarm(data=ReplicateAverages, size=8)
      + stat_compare_means(data=ReplicateAverages, comparisons = list(c("Control", "Drug")),
      method="t.test", paired=TRUE) + theme(legend.position="none")
```

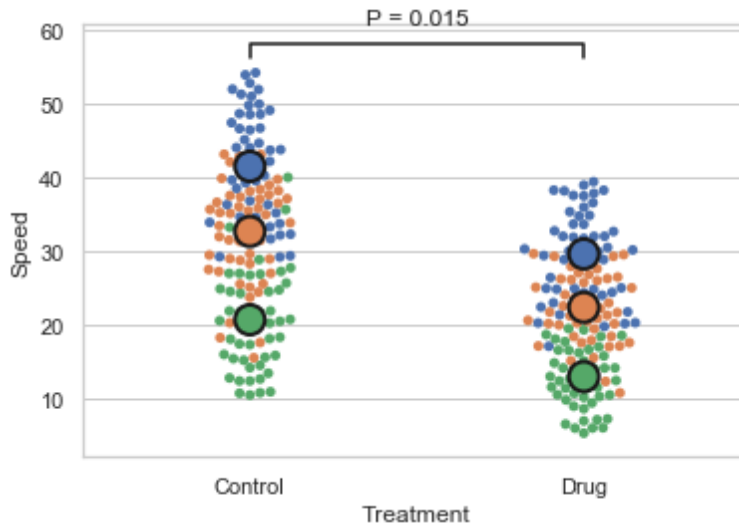

**Figure S5: Tutorial for Making SuperPlots in Python**
Here is some simple code to help make SuperPlots in Python using the Matplotlib, Pandas, Numpy, Scipy, and Seaborn packages. Dataset "combined.csv" is included in the supporting information.

```python
import matplotlib.pyplot as plt
import seaborn as sns
import pandas as pd
import numpy as np
import scipy

#import dataset called "combined" that has the columns "Replicate," "Treatment," and "Speed."
	Change these and the code below if necessary.

combined = pd.read_csv("combined.csv")

sns.set(style="whitegrid")

#calculate the average value for each sample
ReplicateAverages = combined.groupby(['Treatment','Replicate'], as_index=False).agg({'Speed':
	"mean"})

#calculate P value of the sample averages using paired t test
ReplicateAvePivot = ReplicateAverages.pivot_table(columns='Treatment', values='Speed',
	index="Replicate")
statistic, pvalue = scipy.stats.ttest_rel(ReplicateAvePivot['Control'],
	ReplicateAvePivot['Drug'])

P_value = str(float(round(pvalue, 3)))

sns.swarmplot(x="Treatment", y="Speed", hue="Replicate", data=combined)
ax = sns.swarmplot(x="Treatment", y="Speed", hue="Replicate", size=15, edgecolor="k",
	linewidth=2, data=ReplicateAverages)
ax.legend_.remove()
x1, x2 = 0, 1 # columns 'Control' and 'Drug'
y, h, col = combined['Speed'].max() + 2, 2, 'k'
plt.plot([x1, x1, x2, x2], [y, y+h, y+h, y], lw=1.5, c=col)
plt.text((x1+x2)*.5, y+h*2, "P = "+P_value, ha='center', va='bottom', color=col)
```

**Table 1. How the choice of *N* influences conclusions**

| Experiment | If *N* = number of observations | If *N* = number of experiments | Potential Outcomes & problems |
|---|---|---|---|
| A researcher measures the speeds of 20 crawling cells per condition, and repeats the experiment on 3 different days. | *N* = 60 cells<br>Test: ANOVA + Tukey post test<br>*P* = 0.044, *P* = 0.029<br>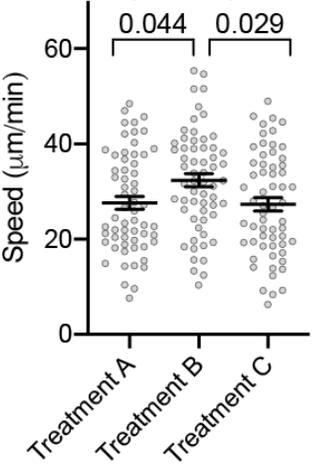 | *N* = 3 experiments<br>Test: ANOVA + Tukey post test<br>*P* = 0.49, *P* = 0.38<br>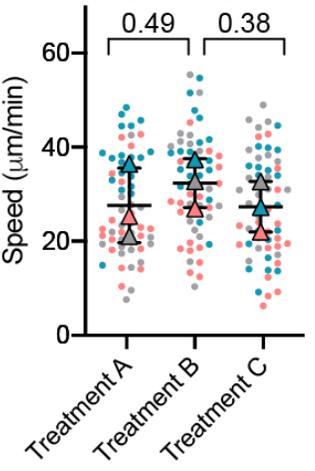 | Here, the researcher would come to a different conclusion based on what they consider "*N*." In this example, data were collected on cells prior to treatment, so they are all untreated. Therefore, differences are due to unpredictable sample-to-sample fluctuations. This example highlights that these chance differences are amplified when each cell is considered its own experiment. When comparing actual treatments, using *N* = 60 could lead to erroneously small *P* values. |
| A researcher measures the pixel intensity of actin staining in cell protrusions. The researcher measures 10 protrusions per cell for 5 cells, and repeats the experiment 3 times. | *N* = 150 protrusions<br>Test: unpaired t-test<br>*P* = 8.9 × 10$^{-33}$<br>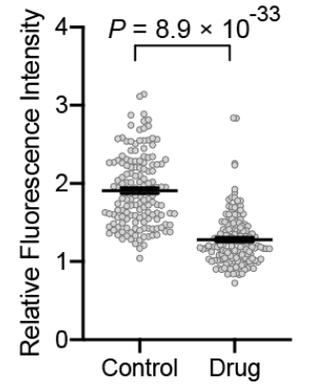 | *N* = 3 experiments<br>Test: paired t-test<br>*P* = 0.041<br>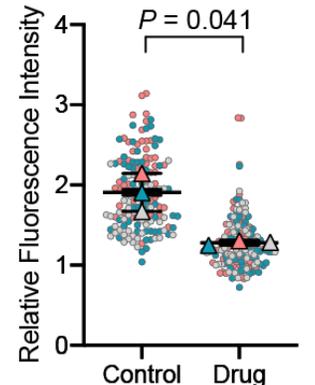 | In this example, there are three choices for N: the number of protrusions, the number of cells, or the number of experiments. In this case (Velle and Campellone, 2018), the *P* value is less than 0.05 regardless of which *N* is used for statistical analysis. This example illustrates that while a miniscule *P* value could raise red flags, the underlying conclusions may still be appropriate.<br><br>While using *N* = 150 may seem preposterous, it's understandable that a researcher measuring >1,800 protrusions by hand would begin to think of each cell as a separate experiment. |

|  | Or, N=15 cells<br>Test: unpaired t-test |  |  |
| --- | --- | --- | --- |
| Researchers compare nucleus size within tumors vs the surrounding cells. They look at 100 transformed and normal cells in 5 different tissue samples. | $N = 500$<br><br>They plot the nucleus size for each cell as a beeswarm plot with tiny error bars and a vanishingly small $P$ value. | $N = 5$<br><br>They compare the nucleus size in transformed and normal cells, paired by tissue sample. Now they have five biological replicates, each encompassing 100 technical replicates. They perform a paired test and get a more reasonable $P$ value. | By using an $N$ of 500, it is likely that the $P$ value will be artificially smaller. By averaging the technical replicates and using an $N$ of 5, the researchers can confirm if the observed difference across patient samples is larger than the natural variation within any single tissue sample. |
| In an in vitro experiment, researchers compare the rates of filament growth of actin orthologs. | $N = 1000$<br><br>They measure the growth rate of thousands of filaments from each organism. They then calculate the statistics using the number of filaments as N, resulting in a tiny $P$ value. | $N = 4$<br><br>They repeat the experiment over multiple days, with different stocks of purified proteins. They calculate the mean filament growth rate in each run and report a $P$ value based on those means. | Counting each filament as a separate sample provides an estimate of the inherent variability of filament growth rate within that one sample. But using that to then compare two orthologs on different coverslips is not appropriate, because there is nothing controlling for different handling of the two proteins during imaging.<br><br>Repeating the experiment with different stocks of protein takes into account random and systematic errors including concentration/pipetting error, room temperature fluctuations, protein degradation, and other, less foreseeable, variables. Readers would want to know if any findings reported can be replicated under similar conditions, for example in another lab. |
| Researchers use laser ablation to cut individual spindle fibers, then observe the downstream effects in those same individual cells. In other cells in the same sample, they apply the same laser dose, but not directed at spindle fibers. | $N = 30$<br><br>They perform the treatment or control on 30 different cells, approaching each cell as an $N$ and calculating a $P$ value to compare the two treatments. |  | Because each cell can be randomly assigned to the control or treated group, it is appropriate in this case that cell is its own sample and calculate a $P$ value from one day's experiment using $N$ as the number of cells. The researchers may conclude that cells treated with laser ablation differ significantly from those that don't have their spindle fibers cut. |
| | | | But that may only be true for that cell strain at that particular passage number and at a specific temperature of the scope room. Observing similar results over multiple days or even with different cell types makes this claim robust. |